\documentclass[a4paper]{article}
\usepackage{amsmath,amssymb}
\usepackage[dvipdfmx]{graphicx}
\usepackage{enumerate}

\newcommand{\e}[1]{\mathrm{e}^{#1}}
\newcommand{\gtatami}[1]{\ensuremath{g(#1)}}

\title{Monte Carlo estimation of the number\\ of tatami tilings}
\author{Kenji Kimura and Saburo Higuchi \\
  Department of Applied Mathematics and Informatics,\\
  Ryukoku University, Otsu, Shiga 520-2194, Japan
}

\begin{document}
\maketitle

\begin{abstract}
Motivated by the way Japanese tatami mats are placed on the floor, 
we consider domino tilings with a constraint and estimate the number of such tilings of plane regions.
We map the system onto a monomer-dimer model with a novel local interaction on the dual lattice. We make use of a variant of the Hamiltonian replica exchange Monte Carlo method where data for ferromagnetic and anti-ferromagnetic models are combined to make a single family of histograms. The properties of the density of states is studied beyond exact enumeration and combinatorial methods. The logarithm of the number of the tilings is linear in the boundary length of the region for all the regions studied.
\end{abstract}

\section{Introduction}
A domino tiling is a non-overwrapping covering of a planar region with rectangular tiles of edge lengths $2\times 1$. It has been a subject of active study in combinatorics and statistical mechanics since the number of domino tilings in  rectangular regions was found\cite{kasteleyn1961statistics,temperley1961dimer}.

One can find an implementation of domino tilings in arrangements of full tatami mats on the floor of old Japanese style rooms. The floor is covered with $1.82\mathrm{m}\times0.91\mathrm{m}$-sized rectangular mats. Traditionally, tatami mats are arranged under the condition that four corners meeting at a point is avoided because the pattern is believed to be inauspicious. In this article, we call this additional rule \textsl{the tatami condition}. 

One can generalize the tiling by allowing $1\times1$ tiles or tatami half-mats. The tatami condition makes sense even if half-mats are present. It states no four corners meet regardless the type of tiles. A tiling with $2\times1$ and $1\times1$ tiles that satisfies the tatami condition on all the vertices shall be called a \textsl{tatami tiling}. 

The rectangle-square tiling model without tatami condition has been studied by mapping it onto the monomer-dimer models. Various interactions has been considered for that model, namely nearest neighbor monomer-monomer, monomer-dimer, dimer-dimer interactions, the one depending on the orientation of dimers, etc. in addition to the hard-core one. The tatami condition is realized as a novel nearest neighbor four-body interaction, which is worth investigation in its own right. As we find below, the tatami condition makes the Monte Carlo sampling hard, it is a suitable model for testing techniques to improve sampling efficiency.

For this model, a natural combinatorial question arises: How many tatami tilings are there for a given region? Erickson \textsl{et.\ al.\ } obtained the number for small rectangular regions in refs.~\cite{ruskey2009counting,erickson2011monomer,mathar2013paving} with the transfer matrix approach. 
Further, they studied the tilings when the number of square tiles is given. Let $t_{\ell_1,\ell_2}(m)$ be the number of tatami tilings with  $m$ squares on $\ell_1\times\ell_2$ rectangular region. They obtained
\begin{equation}
t_{\ell,\ell}(m)=
\begin{cases}
m\cdot2^m+(m+1)\cdot2^{m+1}  & (m<\ell, m \equiv \ell \bmod 2)\\
\ell\cdot2^{\ell-1}  & (m=\ell)\\
0 & (\text{otherwise}).
\end{cases}
\label{square-fixedmonomer}
\end{equation}
via a beautiful combinatorial argument\cite{erickson2012monomer,erickson2013enumerating}.
The number of all the tatami tiling is obtained as
\begin{equation}
\sum_{m=0}^{\ell^2} t_{\ell,\ell}(m)=2^{\ell-1}(3\ell-4)+2. \label{square}
\end{equation}
This expression suggests that the degrees of freedom of this statistical model lives on the boundary because the free energy or the logarithm of \eqref{square} is of order $\mathcal{O}(\ell)$ instead of order $\mathcal{O}(\ell^2)$ \cite{erickson2011monomer}.

The square-rectangle tiling without tatami condition has bulk degrees of freedom: the logarithm of the number of configurations has order $\mathcal{O}(\ell^2)$. Therefore the tatami condition seems to change the nature of the system. It is desirable to study the interpolating systems and find the number of configurations with the number of tatami condition violated points given. It is interesting to ask whether this phenomenon occurs on other geometries. 
Combinatorial methods \cite{erickson2012monomer,erickson2013enumerating} , however,  makes full use of the strong condition: the tatami condition and the geometry. Therefore it is hard to apply to other situations. Methods applicable to general cases are desirable.

The purpose of this work is two-fold. One is to provide an answer to an interesting combinatorial problem, the estimation of the number of configurations of the tatami tilings on a given region. The other is to develop a Monte Carlo (MC) configuration counting technique for this specific hardly relaxing system.

In this article, we generate and count them stochastically by the Markov chain Monte Carlo (MCMC) method in contrast to combinatorial generation of tilings\cite{erickson2014generating}.
It works for arbitrary geometries as well as the simple ones for which exact enumeration or transfer matrix technique applies. 
We employ replica exchange Markov chain Monte Carlo\cite{hukushima1996exchange} to count the frequency of visits to each energy state. Then histogram reweighting \cite{ferrenberg1989optimized,hukushima2002extended} is applied to obtain the density of states.

This article is organized as follows. We define the statistical model in Sec.~2 and describe its exact properties in Sec.~3. The method of MC simulation is explained Sec.~4. We show our estimation in Sec.~5 and state our conclusion in Sec.~6.
 
\section{Model}
Let $R$ be a finite collection of faces $(n,n+1)\times(m,m+1) \subset \mathbb{R}^2$  ($n,m\in\mathbb{Z}$) of the planar square lattice.
A configuration $s$ is an assignment of domino tiles on $R$. 
To be precise, we put $2\times1$-sized domino tiles each of which covers two adjacent faces and does not overwrap with other tiles. All the faces that are not covered with the domino tiles shall be covered with square tiles of $1\times1$ size. 
We say that a configuration $s$ satisfies \emph{the tatami condition} on a vertex when less than four tiles meet there.

\begin{figure}[htbp]
  \includegraphics[width=0.8\linewidth]{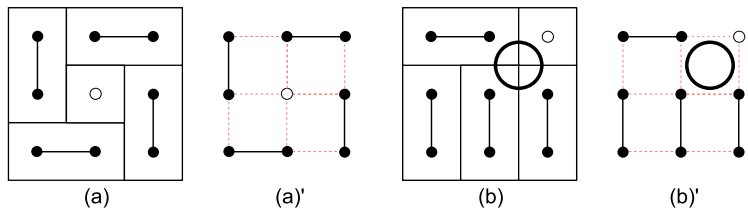}
  \centering
\caption{(a) An example of tatami tilings with four $2\times1$ tiles and a $1\times1$ tile.  (b) A tiling for which the tatami condition is broken at a vertex (indicated by a circle). (a)' and (b)' are the monomer-dimer representation of (a) and (b). \label{fig:monomer-dimer}}
\end{figure}

We map this model onto a monomer-dimer model on the dual square lattice as depicted in Fig.~\ref{fig:monomer-dimer}.
A domino tile corresponds to a dimer which is placed on a dual edge. 
A dual vertex $v^*$ is an endpoint of at most one dimer. 
If no dimer touches $v^*$, one says that a monomer occupies $v^*$.

We consider a monomer-dimer model defined with the partition function
\begin{equation}
  Z(\beta;J_\mathrm{t},J_\mathrm{d})=\sum_{s:\text{monomer-dimer covering}} \e{-\beta H(s;J_\mathrm{t},J_\mathrm{d})}, \label{partition}
\end{equation}
where $\beta$ is the inverse temperature and the Hamiltonian is given by
\begin{equation}
  H(s;J_{\mathrm{t}},J_{\mathrm{d}})= 
J_\mathrm{t} N_\mathrm{t}(s)
+J_\mathrm{d} N_\mathrm{d}(s).
\label{ham}
\end{equation}
The quantities 
\begin{align}
 N_\mathrm{t}(s)=&\sum_{f^*:\text{dual face}} n_\mathrm{t}(f^*,s), \\
 N_\mathrm{d}(s)=&\sum_{v^*:\text{dual vertex}} n_\mathrm{d}(v^*,s)
\end{align}
are the number of dual faces where the tatami condition is broken and that of dimers, respectively. The coefficients $J_\mathrm{t}$ and $J_\mathrm{d}$ can be chosen arbitrarily.
The term $N_\mathrm{t}$ can be regarded as a local multi-tile interaction in the original picture.
Both can be written as sums of local quantities.
\begin{align}
  n_\mathrm{t}(f^*,s)=&
  \begin{cases}
    0 & \text{if some of dual edges around $f^*$ are occupied by dimers}\\
    1 & \text{otherwise}
  \end{cases},\\
  n_\mathrm{d}(v^*,s)=&
  \begin{cases}
   \tfrac12 &  \text{if $v^*$ is occupied by a dimer}\\
   0 &  \text{otherwise}
  \end{cases}.
\end{align}

In this article, we always impose the \emph{hard wall} boundary condition
\footnote{One could  impose an alternative boundary condition, namely the \emph{open} boundary condition, where the compliment of $R$ is populated with $1\times1$ tiles.}, that is, the tatami condition always holds on vertices on the boundary.

The partition function \eqref{partition} can be rewritten as
\begin{equation}
  Z(\beta;J_\mathrm{t},J_\mathrm{d})=\sum_{N,M} g(N,M)\; \e{-\beta (J_\mathrm{t}N+J_\mathrm{d}M)}, \label{density}
\end{equation}
where $g(N,M)$ is the number of configurations with $N$ tatami condition violated dual faces and $M$ dimers.
If we define
\begin{equation}
\gtatami{N}=\sum_{M} g(N,M), \label{gtatami} 
\end{equation}
the number of all the tatami tilings is given by $\gtatami{0}$.

\section{Exact Relations}
Though our method is applicable to arbitrary plane regions, 
in this article, we report the results of rectangular and aztec diamond regions. An aztec diamond of order $\ell$ consists of all faces whose center $(x,y)$ satisfy $|x| + |y| \leq \ell$. 
It consists of  $2\ell(\ell+1)$ faces.
For regions of this shape, the number of tatami tilings is not known exactly but , for small fixed $\ell$, can be counted with an algorithm applicable to arbitrary lattices~\cite{krauth2006algorithms}.

For a fixed region, let $N_\text{max}$ be the largest $N$ such that $g(N)>0$. 
It can be shown that
$N_\text{max}=(\ell_1-1)\times(\ell_2-1)$
for $\ell_1\times \ell_2$ rectangular region, and 
$N_\text{max}=2\ell(\ell-1)+1$ for the aztec diamond of order $\ell$.
For large rectangular or aztec diamond regions, the relation
\begin{equation}
 \gtatami{N_\text{max}}=1 \label{exact}
\end{equation}
holds.  This is because $N_\text{max}$ is attained solely by the state $s$ consisting only of monomers\footnote{Large regions with smooth boundaries have this property. There are a number of regions for which $\gtatami{N_\text{max}}$ is not unity but is known exactly. For example, $\gtatami{N_\text{max}=0}=3$ for $1\times3$ rectangular region. Note the hard-wall boundary condition.}.
This relation will be used for normalizing the density of states from Monte Carlo samples in Sec. \ref{subsec:one-par}. 

For some regions, we have more exact relations for $g$. For example,
we have
\begin{equation}
\gtatami{N_\text{max}-1}=2(\ell_1-1)+2(\ell_2-1) \label{eq:one-dimer}
\end{equation}
for $\ell_1\times \ell_2$ rectangular region.
This is the number of $M=1$ configurations for which the dimer is placed along a boundary. These additional relations is useful for the normalizing the density and for verifying the accuracy of the numerical calculation.

\section{Estimation Method}
We estimate the density of states $g(N,M)$ and $g(N)$ defined in eqs.~\eqref{density} and \eqref{gtatami}.

\subsection{Monte Carlo Moves}
\begin{figure}[htbp]
\begin{center}
\includegraphics[width=0.7\linewidth]{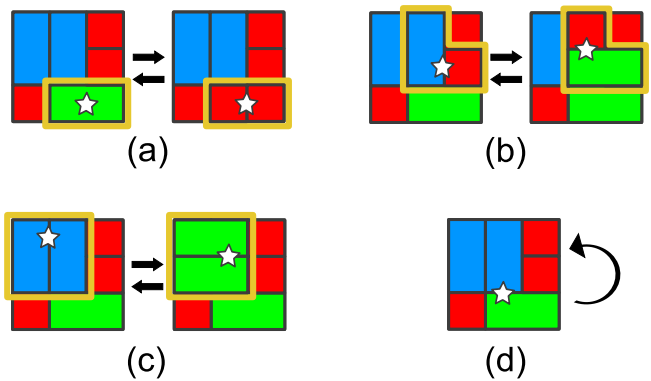}
\end{center}
\caption{Set of moves. (a) If the chosen edge is the center of a dimer or two monomers, change the state to two monomers or a dimer, respectively. (b) If the edge is between a monomer and a dimer, swap them. (c) If the edge is one of two edges shared by two dimers, rotate the dimers $\pi/2$. (d) If none of (a)--(c) applies,  do nothing.}
\label{move}
\end{figure}

We consider a Markov chain on the configuration space $\{s\}$ which satisfies the detailed balance condition and has the distribution \eqref{partition} as the stationary state.
To make a transition, we choose an edge uniformly random and apply the move corresponding to the local configuration as depicted in Fig.~\ref{move}. 
 Then we accept or reject the new configuration using Metropolis algorithm with Hamiltonian \eqref{ham}.
In ref.~\cite{KRS}, the Markov chain with the moves (a) and (b) only was used. We propose to add the move (c) in order to sample configurations with many dimers and many tatami condition violated dual faces efficiently. 

\subsection{Estimation of  $g(N)$}\label{subsec:one-par}
To count the number of tatami configurations $\gtatami{0}$,
we set $J_\mathrm{d}=0$.  We also set $J_\mathrm{t}=+1$ without loss of generality.

We run the simulation at a series of temperatures $\{\beta_j\}$.
Though one could obtain $Z\rightarrow\gtatami{0}$ by sending $\beta$ to infinity, the histogram reweighting\cite{ferrenberg1989optimized,hukushima1999prob} is more efficient. 
Let $h_{N,j}$ be the number of visits to each energy level $N=N_\mathrm{t}$ normalized by all the visits at $\beta_j$. We note that in the limit of infinite sample size, $h_{N,j}$ converges to 
\begin{equation*}
  h_{N,j}=\frac{g(N)\exp(-\beta_j N)}{Z(\beta_j)}
\end{equation*}
due to the law of large numbers. We have 
\begin{equation*}
  g(N)=\frac{h_{N,j}}{\exp(-\beta_j N)/Z(\beta_j)}
\end{equation*}
independent of temperatures once $Z(\beta_j)$ is known.

For finite samples, these estimates have statistical errors.
In ref. \cite{hukushima1999prob}, a single estimate is proposed by composing sums in the numerator and the denominator as
\begin{equation}
  g(N)=\frac{ \sum_{j} h_{N,j}}{\sum_{j} \exp(-\beta_jN)/Z(\beta_j)},
               \label{reweighting1-1}
\end{equation}
where $Z(\beta_j)$ is determined self-consistently by
\begin{equation}
  Z(\beta_j)=\sum_{N} g(N) \exp(-\beta_jN).
               \label{reweighting1-2}
\end{equation}
In ref.\cite{PhysRevLett.98.200601}, it is proved that this set of equations gives the maximum likelihood estimation of $A\cdot\gtatami{N}$ ($N=0,\ldots,N_\text{max}$) up to an over-all factor $A$. In practice, we start with $g(N)=1$ and $Z(\beta_j)=1$ and solve eqs.~\eqref{reweighting1-1},\eqref{reweighting1-2} iteratively for two families of unknowns $g(N)$ and $Z(\beta_j)$. The over-all factor $A$  can be fixed by eq.~\eqref{exact}. 

It is numerically hard to do calibration at $N=N_\mathrm{max}$.
High energy states near $N=N_\text{max}$ are sampled rarely due to the penalty term $J_\mathrm{t}>0$ and small density of states (see eqs.~\eqref{exact},\eqref{eq:one-dimer}).
This can cause serious loss of accuracy of the estimate for $A\cdot \gtatami{N_\text{max}}$.

One would have a large sample if $J_\mathrm{t}$ was negative and the tatami condition breaking vertices were favored. Obviously, this swaps the situations of small and large $N$ regions; negative $J_\mathrm{t}$ makes the estimate of $A\cdot\gtatami{0}$ inaccurate.  A cure is to measure $A_\pm\cdot \gtatami{N}_\pm$ for large (small) $N$ with positive (negative) $J_\mathrm{t}$ respectively and match the estimated values as $\gtatami{N_\mathrm{m}}_+=\gtatami{N_\mathrm{m}}_-$  at some intermediate value $N_\text{m} \simeq N_\text{max}/2$. The problem is that there is no single way to choose $N_\text{m}$. One could even minimize the sum of differences at several $N_\text{m}$'s.

Instead of matching two groups of data as above, we redefine $\tilde{\beta}=\beta\cdot J_\mathrm{t}=\pm \beta$ and consider a series of `temperatures'
\begin{equation}
\tilde{\beta}_j= j\cdot \Delta \beta, \label{posnegtemp}
\end{equation}
where $j$ takes zero, positive and negative integer values.
By running simulations at these `temperatures', a single set of histograms is generated. This method is superior to the former matching method in that it is simply the maximum likelihood estimation and is free of ambiguity of the matching condition. The iterative process takes longer to converge though.

One could say that data for ferromagnetic and anti-ferromagnetic models are combined to make a single family of histograms. We can apply the ordinary reweighting equations \eqref{reweighting1-1},  \eqref{reweighting1-2} to obtain $\gtatami{N}$ because reweighting works not only for thermodynamic distributions but also for general probability distributions\cite{sugita2000multidimensional,fukunishi2002hamiltonian,ejiri2004remarks}.

\subsection{Replica Exchange Monte Carlo}
We speed up the simulation by the replica exchange Monte Carlo\cite{hukushima1996exchange} as was done for configuration counting problem in ref.~\cite{hukushima1999prob}.

For $\tilde{\beta}$ defined in eq.~\eqref{posnegtemp}, we consider exchanges of replica pairs at neighboring temperatures, including the pairs 
$(\tilde{\beta}_j,\tilde{\beta}_{j+1})=(-\Delta\beta,0), (0,+\Delta\beta)$.
The procedure can be regarded as a variant of Hamiltonian replica exchange method\cite{sugita2000multidimensional,fukunishi2002hamiltonian} where one considers replicas having different Hamiltonians. To our knowledge, however, there  has been no application in which Hamiltonians with positive and negative coefficients are exchanged.

\subsection{Estimation of $g(N,M)$}\label{subsec:method-monomer-specified}
To estimate the number of configurations for the number of dimers given,
we make both $J_\mathrm{t}$ and $J_\mathrm{d}$ vary.
In this case, we set $\beta=1$ 
and regard $(J_\mathrm{t},J_\mathrm{d})$ as  free parameters.
We run multi-dimensional replica exchange Monte Carlo\cite{sugita2000multidimensional,fukunishi2002hamiltonian} with replicas having the set of parameters
\begin{equation}
 (J_{\mathrm{t},j},J_{\mathrm{d},k})=(j\cdot\Delta J_\mathrm{t},k\cdot\Delta J_\mathrm{d}), 
\end{equation}
where $j,k$ takes positive and negative integer values in the spirit of \eqref{posnegtemp}.

A replica $(j\cdot\Delta J_\mathrm{t},k\cdot\Delta J_\mathrm{d})$
exchanges its configuration with
$((j\pm1)\cdot\Delta J_\mathrm{t},k\cdot\Delta J_\mathrm{d})$ 
and
$(j\cdot\Delta J_\mathrm{t},(k\pm1)\cdot\Delta J_\mathrm{d})$.

Using multi-parameter reweighting\cite{newman1999monte,fodor2002new,ejiri2004remarks}, we obtain the multi-dimensional density of states $g(N,M)$. Namely, one iteratively solve the following set of equations
\begin{align}
  g(N,M)=&\frac{ \sum_{j,k} h_{NM,jk}}{\sum_{j,k} \exp(-(J_{\mathrm{t},j}N+J_{\mathrm{d},k}M))/Z(J_{\mathrm{t},j},J_{\mathrm{d},k})},
           \label{reweighting2-1}
\\
  Z(J_{\mathrm{t},j},J_{\mathrm{d},k})=&\sum_{N,M} g(N,M) \exp(-(J_{\mathrm{t},j}N+J_{\mathrm{d},k}M)),
           \label{reweighting2-2}
\end{align}
which is the straightforward generalization of eqs.\ref{reweighting1-1},\ref{reweighting1-2}. The two-dimensional histogram $h_{NM,jk}$ stands for the number of visits to $(N_\mathrm{t},N_{\mathrm{d}})=(N,M)$ state by the $(j,k)$-th replica normalized by all its visits.

The over-all factor is fixed using an exact relation like \eqref{exact}.
For example, $g(N_\text{max},0)=1$ can be used for rectangular regions.

\section{Results}
We perform the estimation on rectangular regions up to $\ell_1+ \ell_2\leq256$ and aztec diamond regions up to $\ell\leq48$.
To estimate the statistical error, we divide simulation data into $16$ blocks and calculate the variation among block averages.

Simulations have been performed on a computer with modest power.
\begin{itemize}
\item CPU : Xeon E5-2640 $\times 2$
\item memory : $32$GB
\item OS : Windows 7 Professional
\item compiler : Intel C++ Compiler
\end{itemize}

\subsection{Estimation of $\gtatami{N}$ and the Number of Tatami Tilings $\gtatami{0}$}
\begin{figure}[htbp]
\begin{center}
\includegraphics[width=0.8\linewidth]{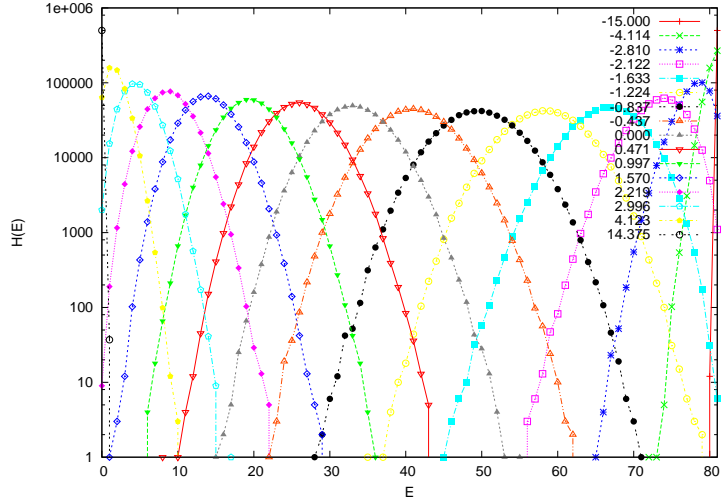}
\end{center}
\caption{Histogram $H(E)$ is $h_{E,j}$ multiplied by the sample size $5\times10^5$ for $10 \times10$  square region. Each curve corresponds to $\tilde{\beta}_j$ shown in the legend.}
\label{his10}
\end{figure}

As an example histogram $h_{N,j}$ for a square region is shown in Fig.~\ref{his10}. One confirms that configuration at all energies $0\leq N \leq (\ell-1)^2$ are visited thanks to the choice \eqref{posnegtemp}.

\begin{figure}[htbp]
\begin{center}
\includegraphics[width=0.8\linewidth]{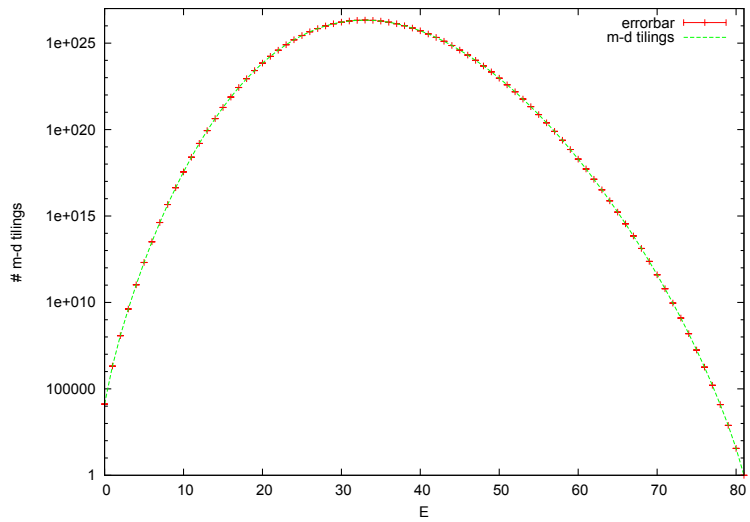}
\end{center}
\caption{Estimation of the number of tilings $\gtatami{E}$ on $10 \times 10$  square region. The over-all factor $A$ is fixed by imposing $g(81)=1$. The number of tatami tilings $g(0)\simeq1.33\times10^4$ can be read on the vertical axis.}
\label{dens}
\end{figure}

\begin{figure}[!p]
\begin{center}
\includegraphics[width=0.8\linewidth]{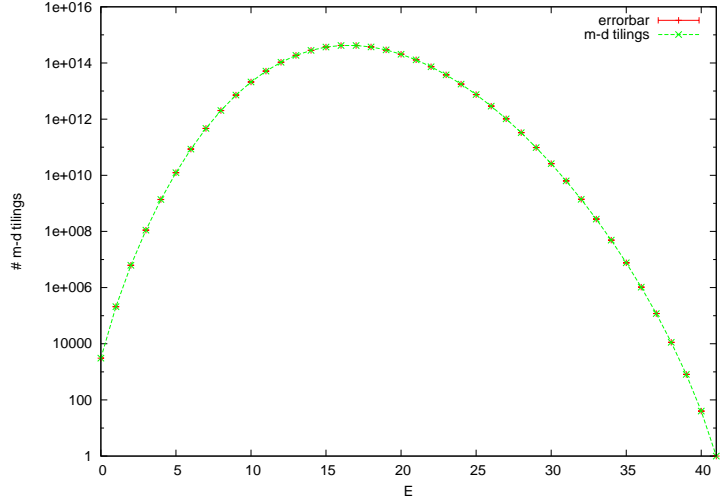}
\end{center}
\caption{Estimation of the number of tilings $\gtatami{E}$ on the aztec diamond region at order $\ell=5$. The over-all factor $A$ is fixed by imposing $g(41)=1$. The number of tatami tilings $g(0)\simeq3.08\times10^3$ can be read on the vertical axis.}
\label{adens}
\end{figure}

The density $\gtatami{N}$ for a square region and that for the aztec diamond region are shown in Figs.~\ref{dens} and \ref{adens}. 
They are all unimodal in $N$ for all the sizes $\ell$ studied.
One finds the estimates for the numbers of tatami tilings as $\gtatami{0}$ in Figs.~\ref{dens} and \ref{adens}.

The estimates $\gtatami{0}$ for square and aztec diamond regions are shown in Tables~\ref{ts} and \ref{ta}, respectively.
They are consistent with the exact result \eqref{square} 
for square regions and the result of direct enumeration for small aztec diamonds within statistical error.

\begin{table}[!p]
\begin{center}
\caption{Number of tatami tilings $g(0)$ of square regions of various sizes. Estimation by MC and exact results are compared.
Statistical error is estimated by dividing the whole simulation into 16 blocks. MCS represents that within one block.}
\begin{tabular}{c r c c}
\hline
Size $\ell\times\ell$ &	Exact\cite{erickson2011monomer} ($2^{\ell-1}(3\ell-4)+2$)& MC result & MCS $\times$ \# of replica\\
\hline
$6\times6$	&	$450$ & $(4.49 \pm 0.03) \times 10^2$ & $(3.0 \times 10^5) \times 48$\\
$7\times7$	& $1090$	 & $(1.09 \pm 0.01) \times 10^3$	&$(3.0 \times 10^5) \times 48$\\
$8\times8$	& $2562$ & $(2.57 \pm 0.02) \times 10^3$ &$(4.0 \times 10^5) \times 48$\\
$9\times9$	& $5890$ &$(5.89 \pm 0.07) \times 10^3$ &$(4.0 \times 10^5) \times 48$\\
$10\times10$	& $13314$& $(1.33 \pm 0.02) \times 10^4$& $ (5.0 \times 10^5) \times 48$\\
$11\times11$	& $29698$ &$(2.96\pm 0.04 ) \times 10^4$ & $ (6.0 \times 10^5) \times 48$\\
$12\times12$	& $65538$ &$(6.56\pm 0.06 ) \times 10^4$ & $ (8.0 \times 10^5) \times 48$\\
$15\times15$	& $671746$ &$(6.70\pm 0.07 ) \times 10^5$ & $ (2.0 \times 10^6) \times 48$\\
$20\times20$	& $29360130$ &$(2.93\pm 0.04 ) \times 10^7$ & $ (5.0 \times 10^6) \times 48$\\
$25\times25$	& $1191182338$ &$(1.19\pm 0.01 ) \times 10^9$ & $ (2.0 \times 10^7) \times 48$\\
$32\times32$	& $197568495618$ &$(2.00\pm 0.07 ) \times 10^{11}$ & $ (1.0 \times 10^6) \times 96$\\
$40\times40$	& $63771674411010$ &$(6.38\pm 0.20 ) \times 10^{13}$ & $ (2.0 \times 10^6) \times 96$\\
$48\times48$	& $1.9703248369746\times 10^{16}$ &$(2.04\pm 0.12 ) \times 10^{16}$ & $ (2.0 \times 10^6) \times 96$\\
$64\times64$	& $1.7339939429287\times 10^{21}$ &$(1.69\pm 0.16 ) \times 10^{21}$ & $ (2.0 \times 10^6) \times 96$\\
$128\times128$	& $6.4653649714978\times 10^{40}$ &$(1.79\pm 1.18 ) \times 10^{41}$ & $ (2.0 \times 10^6) \times 96$\\
\hline
\end{tabular}
\label{ts}
\end{center}
\end{table}

\begin{table}[!p]
\begin{center}
\caption{Number of tatami tilings $g(0)$ of aztec diamond regions of order $\ell$. Estimates by MC and exact results are compared.}
\begin{tabular}{r r c c}
\hline
Order $\ell$ &	Enumeration  & MC & MCS $\times$ \# of replica\\
\hline
$2$	& $80$	&$(8.00 \pm 0.04) \times 10^1$ & $(4.0 \times 10^4) \times 48$\\
$3$	& $392$	&$(3.92 \pm 0.03) \times 10^2$ &$(5.0 \times 10^4) \times 48$\\
$4$	&$1200$& $(1.20 \pm 0.01) \times 10^3$ &$(1.0 \times 10^5) \times 48$\\
$5$	&$3080$& $(3.08 \pm 0.03) \times 10^3$ &$(2.0 \times 10^5) \times 48$\\
$6$	&$7312$& $(7.32 \pm 0.06) \times 10^3$ &$(4.0 \times 10^5) \times 48$\\
$7$	&$16712$& $(1.68 \pm 0.02) \times 10^4$ &$(6.0 \times 10^5) \times 48$\\
$8$	&$37424$& $(3.75 \pm 0.05) \times 10^4$ &$(1.0 \times 10^6) \times 48$\\$10$	&$181392$& $(1.81 \pm 0.02) \times 10^5$ &$(2.0 \times 10^6) \times 48$\\
$12$	&$-$& $(8.51 \pm 0.13) \times 10^5$ &$(2.0 \times 10^6) \times 48$\\
$16$	&$-$& $(1.78 \pm 0.04) \times 10^7$ &$(2.0 \times 10^6) \times 48$\\
$24$	&$-$& $(6.57 \pm 0.34) \times 10^9$ &$(2.0 \times 10^6) \times 48$\\
$32$	&$-$& $(2.30 \pm 0.17) \times 10^{12}$ &$(2.0 \times 10^6) \times 48$\\
$48$	&$-$& $(2.17 \pm 0.27) \times 10^{17}$ &$(2.0 \times 10^6) \times 48$\\
\hline
\end{tabular}
\end{center}
\label{ta}
\end{table}

\subsection{Size Dependence of $g(N)$}
\begin{figure}[!p]
\begin{center}
\includegraphics[width=0.8\linewidth]{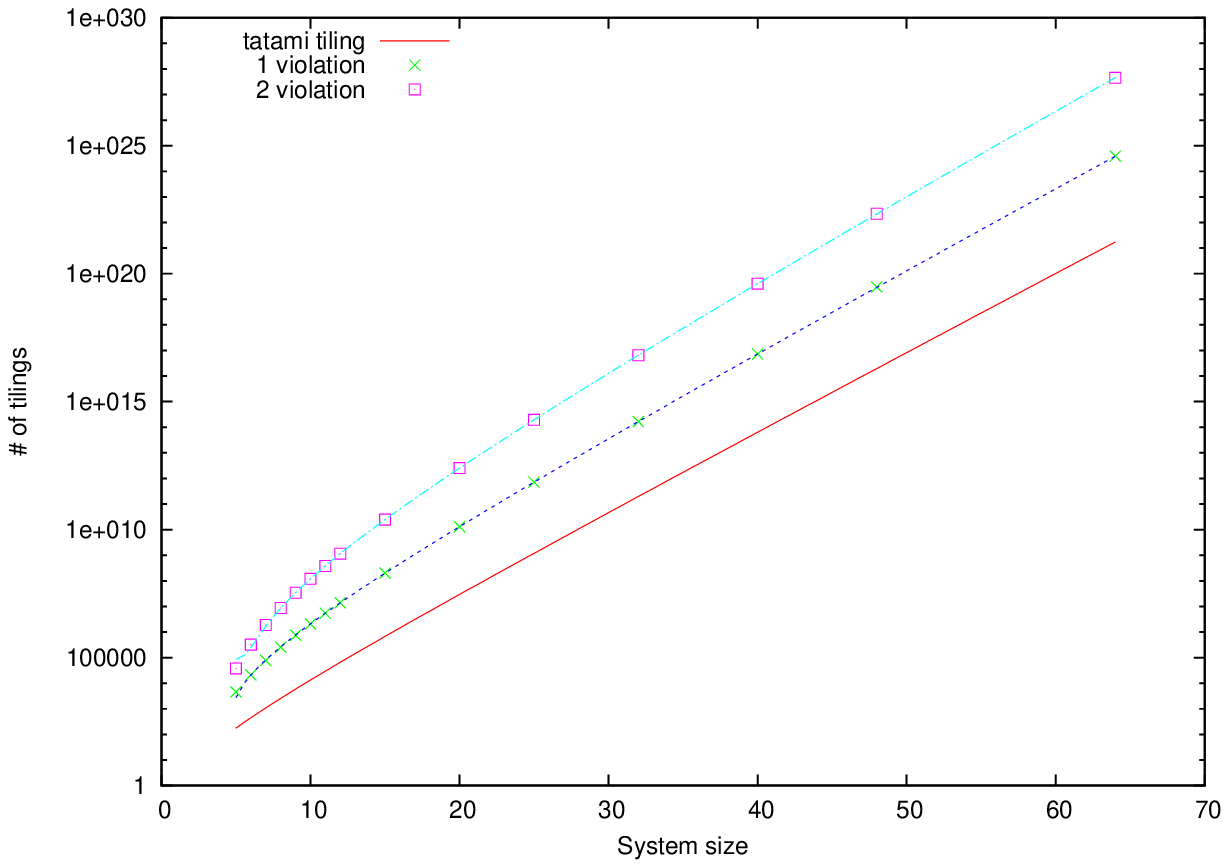}
\end{center}
\caption{The number of tilings with given number of violation. The system size stands for the edge length $\ell$ of square regions.}
\label{violation}
\end{figure}

In contrast to the exact result \eqref{square}, no exact formula for $\gtatami{N}$ are available for $N>0$. We plot $\log \gtatami{N}$ against the size $\ell$ of regions in Fig.~\ref{violation}. It is suggested that $\log \gtatami{N}$  ($N=1,2$) are also of order $\mathcal{O}(\ell)$ in the large $\ell$ limit.

\begin{figure}[!p]
\begin{center}
\includegraphics[width=0.6\linewidth]{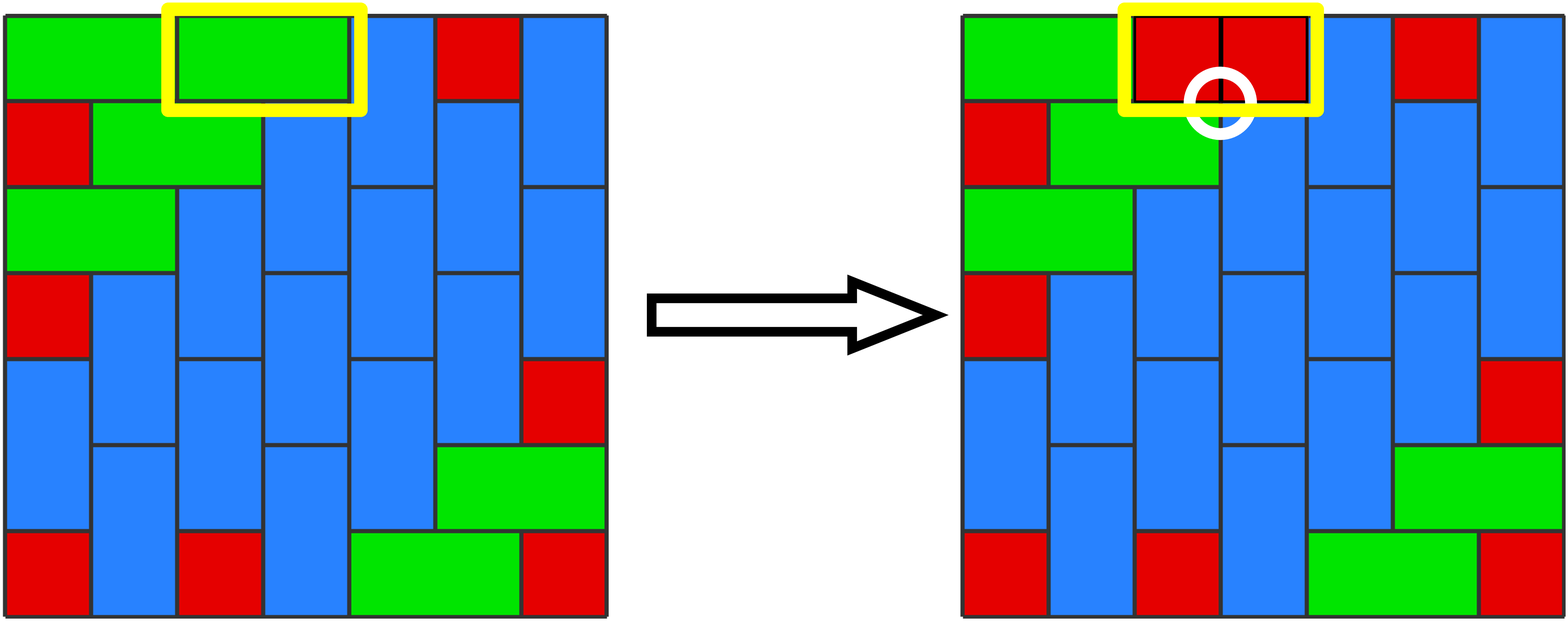}
\end{center}
\caption{An example of obtaining an $N=1$ tiling configuration from $N=0$ one by splitting a dimer into two monomers.}
\label{low-t-e1}
\end{figure}

One can understand this behavior in analogy with the low temperature expansion for spin systems. A generic ground state $N=0$ has $\mathcal{O}(\ell)$ dimers along the boundary. If we split one of them into two monomers, we have an $N=1$ configuration (Fig.~\ref{low-t-e1}). We have more $N=1$ tiling that has a tatami condition violated point in the interior. 
Thus we have $\gtatami{1} \gtrsim \mathcal{O}(g(0))  \times \mathcal{O}({\ell}) $.  

\begin{figure}[!p]
\begin{center}
\includegraphics[width=0.6\linewidth]{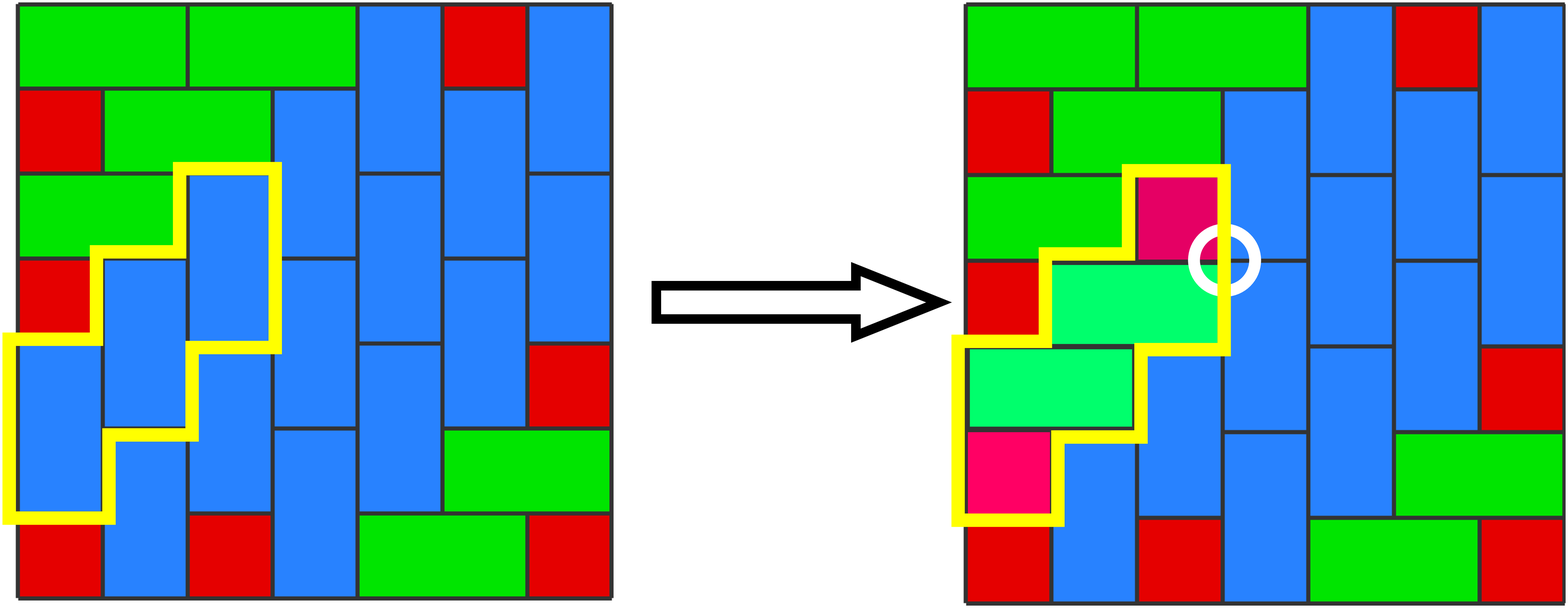}
\end{center}
\caption{An example of $N=1$ configuration that cannot be obtained from $N=0$ one by the splitting procedure in Fig.~\ref{low-t-e1}.}
\label{low-t-e2}
\end{figure}

\subsection{Dependence on Boundary Lengths}
\begin{figure}[!p]
\begin{center}
\includegraphics[width=0.8\linewidth]{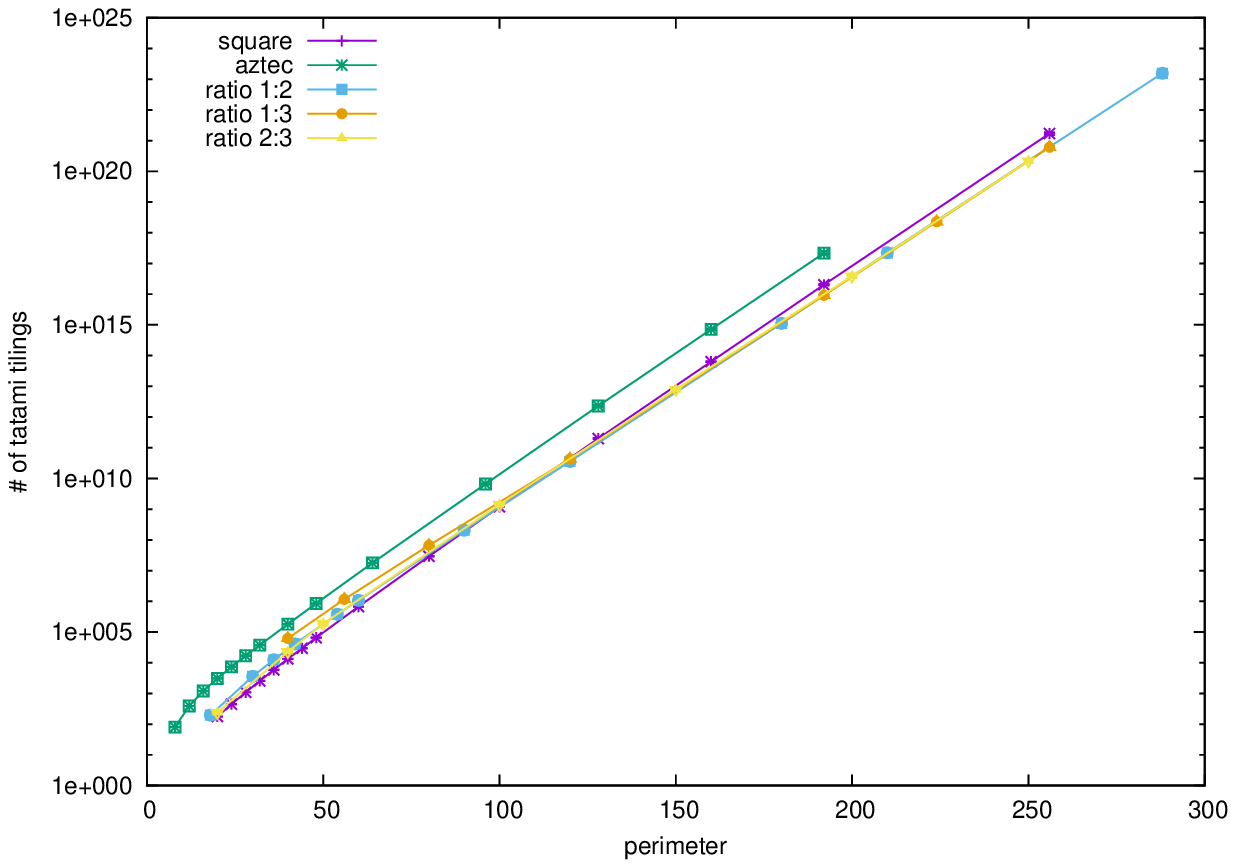}
\end{center}
\caption{The number of tatami tilings $\gtatami{0}$ against boundary length. The vertical axis stands for $\log \gtatami{0}$ while the horizontal axis stands for $p$, the quarter of the boundary length. Ratio 1:2, 1:3 and 2:3 mean rectangles with these aspect ratios.}
\label{perimeter}
\end{figure}

We plot $\log \gtatami{0}$ against the boundary length $p=2(\ell_1+\ell_2)$ for families of rectangles with fixed aspect ratios and $p=4\ell$ for aztec diamonds in Fig.~\ref{perimeter}.
We see that leading contribution to $\log \gtatami{0}$ is of the linear form $C\cdot p+D$. 
When one fits the data with the least square method,
the estimated value of $C$ is within $3$\% of $\frac14\log2$, the exact value \eqref{square} for square regions, for all the families.
On the other hand, the constant term $D$ depends on the shape and the aspect ratio and takes values between $2.0$ and $4.5$.

\subsection{Estimation of $g(N,M)$}
\begin{figure}[!p]
\begin{center}
\includegraphics[width=0.8\linewidth]{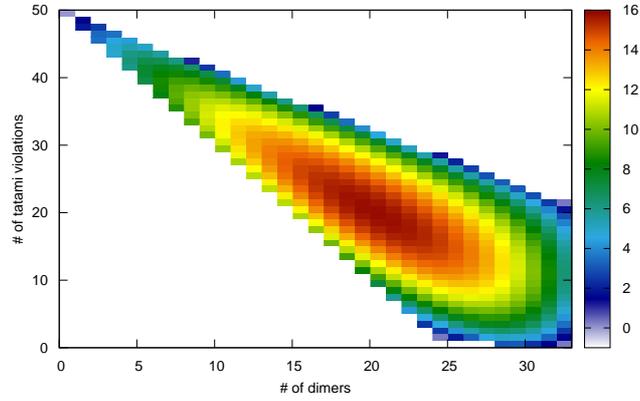}
\end{center}
\caption{Heat map of estimated $\log_{10} g(N,M)$ for the $8 \times 8$ square region. The vertical and the horizontal axes represent $N$ and $M$, respectively.  For white cells, the numbers of tilings are estimated to be zero.}
\label{m-d}
\end{figure}

\begin{figure}[!p]
\begin{center}
\includegraphics[width=0.8\linewidth]{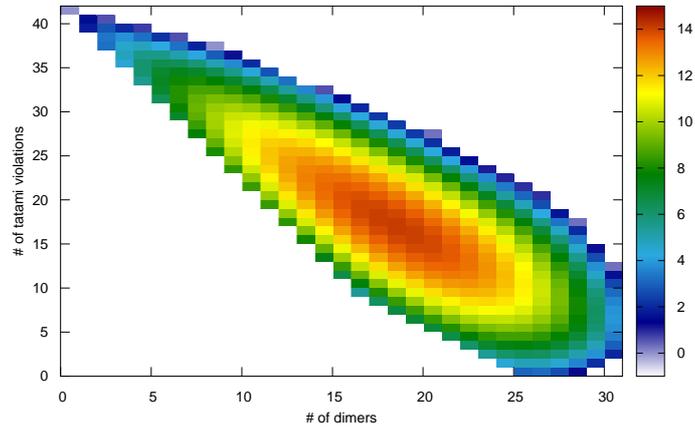}
\end{center}
\caption{Heat map of estimated $\log_{10} g(N,M)$ for the aztec diamond region of order $5$. The vertical and the horizontal axes represent $N$ and $M$, respectively.  For white cells, the numbers of tilings are estimated to be zero.}
\label{m-d-aztec}
\end{figure}

By the method in Sec.~\ref{subsec:method-monomer-specified}, we obtain a table of $g(N,M)$ for each plane region $R$. As examples, data for $8\times8$ square region and the aztec diamond region of order $5$ are drawn as heat maps in Figs.~\ref{m-d} and \ref{m-d-aztec}. 

The fact that an estimated value of $g(N,M)$ is positive implies that the unknown true value is in fact positive. The converse, however, is not always true because the simulation can fail to visit any states at $(N,M)$.

The density profile $g(N,M)$ takes a similar form for all the regions investigated. The density $g$ is non-zero in a domain
\begin{equation*}
 \{(N,M)| N_-(M)\leq N \leq N_+(M)\}.
\end{equation*}
It has a maximum at a cell in the interior of the domain and is decreasing with the distance from that cell.
Two limits $N=N_\pm(M)$ in $(N,M)$ plane cross at $(N,M)=(N_\text{max},0)$.

Then the lower bound is given by $N\geq N_\text{max}-2M$. This inequality can be shown as follows. Consider the tiling $(N,M)=(N_\text{max}, 0)$ consisting only of monomers. Then one introduces dimers one by one. Each dimer creates at most two dual faces where the tatami condition is met. This bound is saturated for small $M$ for the cases investigated.

The upper bound is given by $N\leq N_\text{max}-M$ for small $M$. This comes from the fact that introduction of the first dimer creates, even if it is placed along a boundary line, at least one dual face where the tatami condition is met. For aztec regions it holds only for $M=0,1$ while for rectangular regions it holds for $0\leq M<\min(\ell_1,\ell_2)$.
\nopagebreak

There is a difference between the shapes of the domains for rectangles and those for aztec diamonds, especially near the bottom row $N=0$ which corresponds to tatami tilings. For square region case, the result for the row is consistent with \eqref{square-fixedmonomer}. For aztec diamond region case, the data for the row is consistent with the result of direct enumeration; at order $5$, the tatami tilings with less than two monomers are absent.

\section{Conclusion}
We have studied a monomer-dimer model with an interaction term which enforces the tatami condition. We have been able to estimate the number of the tatami tilings and find the size dependence. Moreover, by two-dimensional replica exchange Monte Carlo and multi-parameter histogram reweighting, we have been able to estimate the number of tilings with specified numbers of dimers and tatami condition violated dual faces. To this end, we have proposed a Monte Carlo method to calculate the density of states of combinatorial models by combining ferromagnetic and anti-ferromagnetic models. Efficiency of this method when applied to other combinatorial problem is left for future study.

\section*{Acknowledgments}
We are grateful to Professors Shinji Iida and Junta Matsukidaira for useful discussions. 

\clearpage

\bibliographystyle{unsrt}  
\bibliography{tatami,mc,dimer}

\begin{thebibliography}{10}

\bibitem{kasteleyn1961statistics}
Pieter~W Kasteleyn.
\newblock The statistics of dimers on a lattice: I. the number of dimer
  arrangements on a quadratic lattice.
\newblock {\em Physica}, 27(12):1209--1225, 1961.

\bibitem{temperley1961dimer}
HNV Temperley and Michael~E Fisher.
\newblock Dimer problem in statistical mechanics-an exact result.
\newblock {\em Philosophical Magazine}, 6(68):1061--1063, 1961.

\bibitem{ruskey2009counting}
Frank Ruskey and Jennifer Woodcock.
\newblock Counting fixed-height tatami tilings.
\newblock {\em the electronic journal of combinatorics}, 16(1):R126, 2009.

\bibitem{erickson2011monomer}
Alejandro Erickson, Frank Ruskey, Jennifer Woodcock, and Mark Schurch.
\newblock Monomer-dimer tatami tilings of rectangular regions.
\newblock {\em the electronic journal of combinatorics}, 18(1):P109, 2011.

\bibitem{mathar2013paving}
Richard~J Mathar.
\newblock Paving rectangular regions with rectangular tiles: tatami and
  non-tatami tilings.
\newblock {\em arXiv preprint arXiv:1311.6135}, 2013.

\bibitem{erickson2012monomer}
Alejandro Erickson and Mark Schurch.
\newblock Monomer-dimer tatami tilings of square regions.
\newblock {\em Journal of Discrete Algorithms}, 16:258--269, 2012.

\bibitem{erickson2013enumerating}
Alejandro Erickson and Frank Ruskey.
\newblock Enumerating maximal tatami mat coverings of square grids with $ v $
  vertical dominoes.
\newblock {\em arXiv preprint arXiv:1304.0070}, 2013.

\bibitem{erickson2014generating}
Alejandro Erickson and Frank Ruskey.
\newblock Generating tatami coverings efficiently.
\newblock {\em arXiv preprint arXiv:1403.4776}, 2014.

\bibitem{hukushima1996exchange}
Koji Hukushima and Koji Nemoto.
\newblock Exchange monte carlo method and application to spin glass
  simulations.
\newblock {\em Journal of the Physical Society of Japan}, 65(6):1604--1608,
  1996.

\bibitem{ferrenberg1989optimized}
Alan~M Ferrenberg and Robert~H Swendsen.
\newblock Optimized monte carlo data analysis.
\newblock {\em Physical Review Letters}, 63(12):1195, 1989.

\bibitem{hukushima2002extended}
Koji Hukushima.
\newblock Extended ensemble monte carlo approach to hardly relaxing problems.
\newblock {\em Computer Physics Communications}, 147(1):77--82, 2002.

\bibitem{krauth2006algorithms}
Werner Krauth.
\newblock Algorithms and computations, 2006.

\bibitem{KRS}
Claire Kenyon, Dana Randall, and Alistair Sinclair.
\newblock Approximating the number of monomer-dimer covering of a lattice.
\newblock {\em Journal of Statistical Physics}, 83:637--659, May 1996.

\bibitem{hukushima1999prob}
Koji Hukushima.
\newblock Domain-wall free energy of spin-glass models: Numerical method and
  boundary conditions.
\newblock {\em Phys. Rev. E}, 60(4):3606--3613, 1999.

\bibitem{PhysRevLett.98.200601}
Michael Habeck.
\newblock Bayesian reconstruction of the density of states.
\newblock {\em Phys. Rev. Lett.}, 98:200601, May 2007.

\bibitem{sugita2000multidimensional}
Yuji Sugita, Akio Kitao, and Yuko Okamoto.
\newblock Multidimensional replica-exchange method for free-energy
  calculations.
\newblock {\em The Journal of Chemical Physics}, 113(15):6042--6051, 2000.

\bibitem{fukunishi2002hamiltonian}
Hiroaki Fukunishi, Osamu Watanabe, and Shoji Takada.
\newblock On the hamiltonian replica exchange method for efficient sampling of
  biomolecular systems: application to protein structure prediction.
\newblock {\em The Journal of Chemical Physics}, 116(20):9058--9067, 2002.

\bibitem{ejiri2004remarks}
Shinji Ejiri.
\newblock Remarks on the multiparameter reweighting method for the study of
  lattice qcd at nonzero temperature and density.
\newblock {\em Physical Review D}, 69(9):094506, 2004.

\bibitem{newman1999monte}
Mark E~J Newman and Gerard~T Barkema.
\newblock {\em Monte Carlo methods in statistical physics}.
\newblock Clarendon Press Oxford, 1999.

\bibitem{fodor2002new}
Zolt{\'a}n Fodor and S{\'a}ndor~D Katz.
\newblock A new method to study lattice qcd at finite temperature and chemical
  potential.
\newblock {\em Physics Letters B}, 534(1):87--92, 2002.

\end{thebibliography}
\end{document}